\documentclass[a4paper,11pt]{article}
\usepackage{jinstpub} 
\usepackage{graphicx}   
\usepackage{placeins}   

\proceeding{Pixel2024: 11$^{\text{th}}$ International Workshop on Semiconductor Pixel Detectors for Particles and Imaging  \\
18-22 Nov 2024\\
Strasbourg (France)}

\title{A Packaging Method for ALPIDE Integration Enabling Flexible and Low-Material-Budget Designs}

\author[a]{D. Novel}

\author[a,b,c,1]{, A. Lega \note{Corresponding author.}}

\author[a]{, T. Facchinelli}

\author[b,c]{, R. Iuppa}

\author[d,e]{, S. Beolé}

\author[a]{, P. Bellutti}

\affiliation[a]{Fondazione Bruno Kessler,\\ 
                 Via Sommarive, 18,
                 38123, 
                 Povo, TN, 
                 Italy}
                 
\affiliation[b]{University of Trento, Physics Department,\\
                 Via Sommarive, 14,
                 38123, 
                 Povo, TN, 
                 Italy}
            
\affiliation[c]{INFN TIFPA,\\ 
                 Via Sommarive, 14,
                 38123, 
                 Povo, TN, 
                 Italy}
                 
\affiliation[d]{University of Turin, \\
Physics Department,
                 Via Pietro Giuria, 1,
                 10125, 
                 Torino, TO, 
                 Italy}
                 
\affiliation[e]{INFN Turin,\\
                 Via Pietro Giuria, 1,
                 10125, 
                 Torino, TO, 
                 Italy}

\emailAdd{alega@fbk.eu}

\abstract{This work presents a novel solution for the packaging of ALPIDE chips that facilitates non-planar assembly with a minimal material budget. This solution represents a technological advancement based on methodologies developed for the ALICE ITS1 and the STAR tracker two decades ago. The core of this approach involves the use of flexible cables composed of aluminum and polyimide, with thicknesses on the order of tens of micrometers. These cables are connected to the sensors using single-point Tape Automated Bonding (spTAB), which replaces the traditional wire bonding technique that is suboptimal for curved integrations. The spTAB bonding is achieved by creating openings in the polyimide layer, allowing aluminum wires to remain free-standing, which are then connected to the sensor using pressure and ultrasonic energy. Extending this concept, we have applied this approach to entire printed circuit boards (PCBs), resulting in a fully flexible packaging solution maintaining an ultra-low material budget. This work introduces a prototype utilizing this method to bond an ALPIDE chip, proposing it as a viable option for future designs necessitating flexible packaging for both the chip and associated electronics. The overall workflow, comprising microfabrication and assembly, is carried out at the Fondazione Bruno Kessler and INFN TIFPA laboratories and will be detailed to elucidate our procedures and demonstrate the applicability of our solution in future experimental setups. The proposed packaging features a flexible PCB constructed from three stacked layers, each containing 20 µm thick aluminum features and a 25 µm thick polyimide substrate. These layers include a ground layer, a signal layer (encompassing both digital and analog signals), and a local bonding layer (which substitutes wire bonding). The spTAB technique is employed for inter-layer connections within the PCB and for sensor bonding. We will discuss the performance of transferring both digital and analog electrical information through the flexible PCBs.}


\begin{document}
\maketitle
\flushbottom

\section{The Need for Flexible, Low-Material-Budget Detector Packaging}

Silicon detector technology represents the state of the art in particle tracking for high-energy physics experiments~\cite{Art:Wermes}, offering exceptional spatial resolution, fast response times, and strong radiation hardness. However, the main contributors to the material budget in modern detectors, such as those in the ALICE ITS2 upgrade~\cite{Art:Felix}, widely considered state of the art in minimizing material budget, are the integration electronics (PCBs) and mechanical supports, rather than the silicon itself~\cite{Art:ITS2}. The choice between copper and aluminum for the PCB material plays a decisive role in the total material budget: copper shorter radiation length increases multiple scattering more than aluminum. In response, emerging detector designs aim to minimize external materials around the sensor. One notable approach is \emph{stitching}, used for future upgrades such as ALICE ITS3~\cite{Art:Stitching}, which allows larger silicon sensors without extra interconnect structures. These developments drastically reduce the material budget and improve performance. Nonetheless, entirely removing support materials remains challenging for many experiments and applications, so lightweight PCB solutions are still critical. This work presents research on developing an innovative packaging system that reduces the material budget, particularly by adopting thin flexible PCBs, while preserving the structural stability required for high-performance silicon detector systems.

\section{Microfabrication of Flexible PCBs}
\FloatBarrier
\begin{figure}[]
    \centering
    \includegraphics[width=1\linewidth]{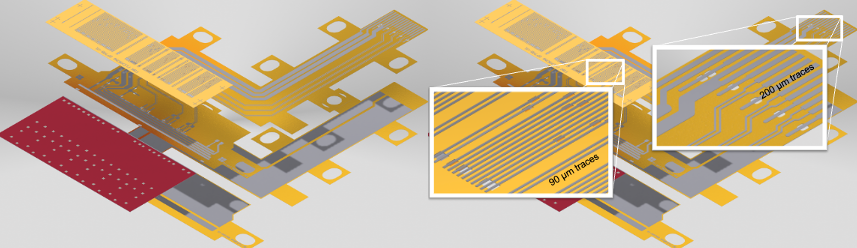}
    \caption{Left: 3D CAD design of the microfabricated PCB developed to interface with the ALPIDE chip. The bonding layer on the top is where spTAB connections are performed, while the flex layer in the middle transmits analog power and digital signals. The bottom layer is the ground (analog and digital). Each single layer is composed of 25~$\mu$m-thick Kapton and 20~$\mu$m-thick aluminum. This design was inspired by an R\&D effort for ITS3~\cite{Art:Langoy} and adapted to accommodate TAB bonding technology and aluminum PCBs. Right: A closer view of the layout adaptation for spTAB.}
    \label{fig:cad_flex}
\end{figure}

The ALICE-ITS1 detector introduced a low-material-budget packaging approach employing ultra-thin cables to connect sensors with front-end electronics~\cite{Art:LTU}. Unlike traditional copper-based PCBs ($X_{0,\text{Cu}} = 1.4\,\text{cm}$), ALICE-ITS1 relied on aluminum ($X_{0,\text{Al}} = 8.9\,\text{cm}$) to reduce multiple scattering and improve tracking precision. The conductor was deposited on Kapton, a flexible polyimide with thicknesses on the order of tens of micrometers, further lowering the material contribution. This single-layer PCB cable enabled reliable connections while minimally disturbing particle trajectories, making ALICE-ITS1 one of the most efficient tracking detectors of its time~\cite{Art:ITS1}. A key innovation was \emph{single point Tape Automated Bonding (spTAB)}~\cite{Art:LTU}, which replaced traditional wire bonding with a more compact method that improved mechanical reliability  \cite{Art:TABvsWIRE}.  \newpage

\noindent Building on ALICE-ITS1, the present work proposes several innovations:
\begin{itemize}
    \item \textbf{Wafer-level PCB production processes}, enabling uniform, large-scale fabrication.
    \item \textbf{Compatibility with silicon cleanrooms}, allowing seamless integration into standard detector-fabrication workflows.
    \item \textbf{A KiCad-based design approach}, ensuring an open-source, highly adaptable layout environment.
\end{itemize}

\noindent By scaling these concepts and integrating them into standard production lines, we aim to develop lightweight, mechanically stable PCBs suited to modern particle detector designs. Such efforts reduce the overall material budget while offering a more accessible path toward next-generation particle detector technologies.

\section{ALPIDE Integration on a Flexible PCB}
\FloatBarrier
\begin{figure}
    \centering
    \includegraphics[width=0.8\linewidth]{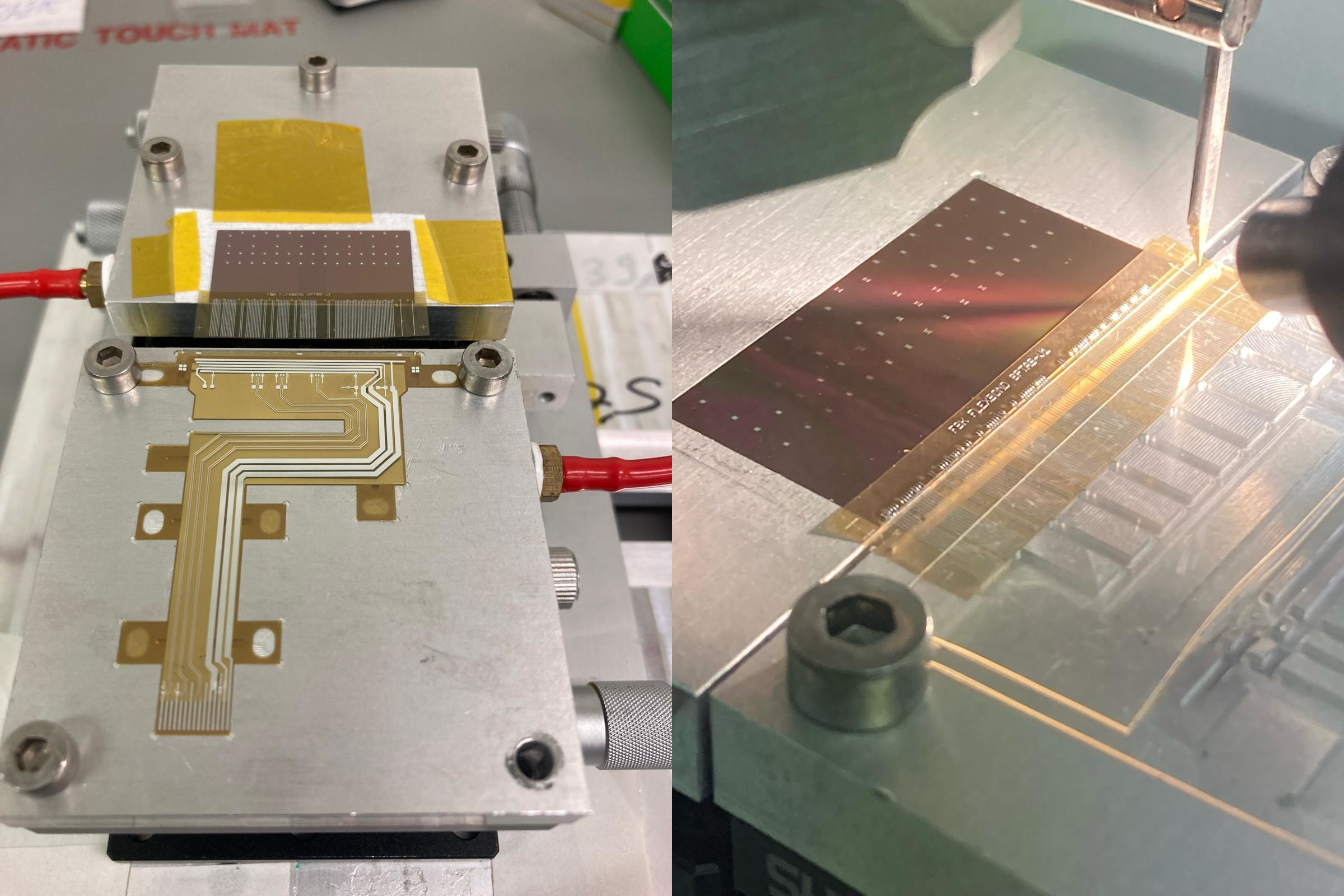}
    \caption{Left: the jig holding the ALPIDE chip and the flexible PCB in two separate sections. Right: the chip bonded to the bonding tape. }
    \label{fig:TAB_jig_2}
\end{figure}

The ALPIDE (ALICE Pixel Detector)~\cite{Art:ALPIDE} is a high-performance monolithic active pixel sensor (MAPS) developed for the Inner Tracking System (ITS) upgrade in ALICE at CERN~\cite{Art:ITS2}. Built using a 180\,nm CMOS  process, it integrates both the sensing element and readout electronics on a single wafer. In the ITS upgrade, ALPIDE 50\,$\mu$m thickness and high spatial resolution significantly enhance tracking precision~\cite{Art:ITS2}. Leveraging the microfabrication method developed at FBK for low-material-budget PCBs~\cite{Art:FBK}, this section details the production of an initial PCB designed to evaluate the overall performance of the fabrication process. Beyond serving as a test structure for process quality, this PCB also validates the overall bonding procedure, confirming its capability to meet the interconnection requirements of a state-of-the-art chip.
\noindent To meet these goals, the PCB was integrated with an ALPIDE chip. A proof-of-concept design was adopted from previous ALICE collaboration work~\cite{Art:Langoy}, then adapted to align with our specific integration techniques (see Fig.~\ref{fig:cad_flex}). The layout was first created in \emph{KiCad}\footnote{\href{https://www.kicad.org}{KiCad}}, chosen for its open-source and user-friendly PCB design interface. For compatibility with FBK microfabrication processes, the layout was then converted into \emph{KLayout}\footnote{\href{https://www.klayout.de}{KLayout}}, enabling the creation of micrometer-scale features. A series of lithography masks was subsequently produced to imprint the designed features onto the kapton-aluminum substrates. One key modification focused on ensuring spTAB compatibility by optimizing the Kapton openings to keep aluminum trace deformation within acceptable limits, resulting in robust and reliable TAB connections. Briefly, the design approach can be summarized by:

\begin{itemize}
    \item \textbf{Bonding on small pad sizes}: $92 \times 92\,\mu\mathrm{m}^2$ (see~\cite{Art:IEEE2} for further details).
    \item \textbf{Three-layer PCB}: Comprising a signal plane, a bonding plane, and a grounding plane (see Fig.~\ref{fig:cad_flex}).
    \item \textbf{Inner barrel mode connection}\footnote{\href{https://sunba2.ba.infn.it/MOSAIC/ALICE-ITS/Documents/ALPIDE-operations-manual-version-0_3.pdf}{ALPIDE Operations Manual}}: All CHIPID pads are left unconnected, resulting in a CHIPID of 0x0.
    \item \textbf{Readout interface}: Pogo pins, flexible printed-circuit connectors, and zero-insertion-force (ZIF) interfaces were adopted to enable FPGA communications (see~\cite{Art:IEEE1} for further details).
\end{itemize}

\noindent In accommodating spTAB, the Kapton openings were sized so that the bonding tip could reliably connect to the ALPIDE chip beneath (Fig.~\ref{fig:TAB_jig_2}). Since each ALPIDE bond pad measures $92 \times 92\,\mu\mathrm{m}^2$, the aluminum traces were designed with matching dimensions. Traces with an actual width of about $70\,\mu\mathrm{m}$ were obtained through isotropic wet-etch. A bonding tip of suitable size is therefore needed for successful interconnection. Moreover, the Kapton-window geometry was chosen to keep the aluminum bending angle below the 15$^\circ$ fracture threshold, leading to a safe $\sim$45$^\circ$ bond angle. The bonding window length was set to $350\,\mu\mathrm{m}$, while its width was matched to the pad dimensions. Based on IPC-2251, the characteristic impedance of the differential traces is approximately $50\,\Omega$, deviating from the $100\,\Omega$ target of the M-LVDS TIA-EIA-899 technology\footnote{\href{https://www.ti.com/lit/an/slla108a/slla108a.pdf}{Texas Instruments Application Report SLLA108A}} used in the ALPIDE chip. Although this mismatch likely may cause reflections and degrade signal integrity at high data rates, it was considered acceptable at this proof-of-concept stage. Future research will optimize these parameters. A dry-etch process was preferred over traditional wet etching for Kapton removal, providing more precise feature control and minimizing undercut~\cite{Art:GEM}. After verifying that the bonded chips were undamaged by connecting nonfunctional samples, the project advanced to functional testing to confirm full ALPIDE operation. An FPGA-based system was used to enable the chip clock and perform register read/write operations. The FPGA architecture used to interface with the chip was adapted from work carried out for the HEPD-02 payload \cite{Art:HEPD02}, enabling straightforward tests on the chip. Both analog and digital power supplies were set to 1.8\,V, and the current measurements indicated correct chip response (30\,mA with the clock off and roughly double with the clock on). Additionally, register write/read tests were conducted, confirming that the chip accurately interprets commands and responds with valid data. These findings indicate that the entire procedure is stable, enabling us to undertake a more detailed, quantitative evaluation of the flexible PCB performance under realistic operating conditions.

\section{Conclusions}
The primary objective of this R\&D was to validate the proposed technology and show that the bonded ALPIDE chips work correctly when bonded to our flexible PCBs. Full chip readout and advanced firmware development for the FPGA will be undertaken in the next phase of the project. Although these preliminary findings are encouraging, further work is required to fully optimize signal transmission through the flexible PCB. In particular, mitigating impedance mismatches and refining the overall design to ensure robust signal integrity may involve adjusting trace dimensions and PCB thicknesses. Future efforts will also include detailed signal integrity analysis, such as eye-diagram testing, to assess how the flexible PCB affects signal quality, alongside thermal and mechanical stress tests to guarantee reliable performance in operational environments. While these steps lie beyond the scope of the current work, they are essential to meeting the project broader objectives and will be pursued in the coming years.

\acknowledgments
The authors gratefully acknowledge Fondazione Caritro for funding this research through the Flexbond Project (Grant No. 10917). They also extend their gratitude to the Center for Sensors and Devices at Fondazione Bruno Kessler, whose cleanroom facilities and technical support were vital to the successful completion of the microfabrication processes. Additionally, the authors thank INFN TIFPA for providing support with the integration of the ALPIDE chip that greatly contributed to this work.



\begin{thebibliography}{99}

\bibitem{Art:Wermes}
Kolanoski, Hermann, and Norbert Wermes, \textit{Particle Detectors: Fundamentals and Applications.} \href{https://books.google.it/books?hl=it&lr=&id=2TjpDwAAQBAJ&oi=fnd&pg=PP1&dq=wermes+norbert+book&ots=CQDeviSqaM&sig=HY5DngA9nRtaP01IA57djm-hqKo&redir_esc=y#v=onepage&q=wermes%20norbert%20book&f=false}{Oxford University Press, USA, 2020}

\bibitem{Art:Felix}
Reidt, Felix, and Alice Collaboration, \textit{Upgrade of the ALICE ITS detector.} \href{https://doi.org/10.1016/j.nima.2022.166632}{Nuclear Instruments and Methods in Physics Research Section A: Accelerators, Spectrometers, Detectors and Associated Equipment 1032, 166632, 2022}

\bibitem{Art:ITS2}
ALICE collaboration, \textit{ALICE upgrades during the LHC Long Shutdown 2.} \href{https://doi.org/10.48550/arXiv.2302.01238}{arXiv preprint arXiv:2302.01238 (2023)}

\bibitem{Art:Stitching}
Rinella, G. Aglieri, and Alice Collaboration, \textit{Developments of stitched monolithic pixel sensors towards the ALICE ITS3.} \href{https://doi.org/10.1016/j.nima.2023.168018}{Nuclear Instruments and Methods in Physics Research Section A: Accelerators, Spectrometers, Detectors and Associated Equipment 1049 (2023): 168018.}

\bibitem{Art:LTU}
Oinonen, Markku, et al., \textit{ALICE silicon strip detector module assembly with single-point TAB interconnections.} \href{https://cds.cern.ch/record/920152/files/p92.pdf}{11th Workshop on Electronics for LHC and Future Experiments, pp.92-97.}

\bibitem{Art:ITS1}
Rashevsky, A., et al., \textit{Large area silicon drift detector for the ALICE experiment.} \href{https://doi.org/10.1016/S0168-9002(02)00531-4}{Nuclear Instruments and Methods in Physics Research Section A: Accelerators, Spectrometers, Detectors and Associated Equipment 485.1-2 (2002): 54-60.}

\bibitem{Art:TABvsWIRE}
Howard, Randy, and John Lynch, \textit{Tape Automated Bonding.} \href{https://www.sae.org/publications/technical-papers/content/871568/}{No. 871568. SAE Technical Paper, 1987.}

\bibitem{Art:ALPIDE}
Mager, M., and ALICE collaboration., \textit{ALPIDE, the Monolithic Active Pixel Sensor for the ALICE ITS upgrade.} \href{https://doi.org/10.1016/j.nima.2015.09.057}{Nuclear Instruments and Methods in Physics Research Section A: Accelerators, Spectrometers, Detectors and Associated Equipment 824 (2016): 434-438.}

\bibitem{Art:FBK}
Novel, D., et al., \textit{Evolution of flexible PCBs in particle detection: From ALICE ITS1 to future frontiers in microfabrication for ALPIDE chip integration.} \href{https://doi.org/10.1016/j.nima.2024.169840}{Nuclear Instruments and Methods in Physics Research Section A: Accelerators, Spectrometers, Detectors and Associated Equipment 1069 (2024): 169840.}

\bibitem{Art:Langoy}
Langøy, Rune, et al., \textit{Thinning and readout during bending of a custom silicon IC.} \href{https://doi.org/10.1109/ESTC48849.2020.9229845}{2020 IEEE 8th Electronics System-Integration Technology Conference (ESTC). IEEE, 2020.}

\bibitem{Art:IEEE2}
Novel, David, et al., \textit{Tape Automated Bonding of an ALPIDE Chip with Flexible PCBs.} \href{https://doi.org/10.1109/ESTC60143.2024.10712095}{2024 IEEE 10th Electronics System-Integration Technology Conference (ESTC). IEEE, 2024.}

\bibitem{Art:IEEE1}
Novel, David, et al., \textit{Electrical Characterization of an ALPIDE Chip TAB Bonded with Flexible PCBs.} \href{https://doi.org/10.1109/ESTC60143.2024.10712103}{2024 IEEE 10th Electronics System-Integration Technology Conference (ESTC). IEEE, 2024.}

\bibitem{Art:GEM}
Lega, A., et al., \textit{Plasma-Based Etching Approach for GEM Detector Microfabrication at FBK for X-ray polarimetry in space.} \href{https://iopscience.iop.org/article/10.1088/1748-0221/19/04/C04017/meta}{Journal of Instrumentation 19.04 (2024): C04017.}


\bibitem{Art:HEPD02}
Nicolaidis, Riccardo, et al., \textit{The TDAQ system of the HEPD-02 on the CSES-02 mission.} \href{https://pos.sissa.it/444/1321/pdf}{POS PROCEEDINGS OF SCIENCE 1321 (2022).}

\end{thebibliography}


\end{document}